\RequirePackage[l2tabu,orthodox]{nag}
\documentclass{article}
\usepackage[affil-it]{authblk}
\pdfoutput=1
\usepackage{amsmath}
\usepackage{amssymb}
\usepackage{graphicx}
\usepackage{subfigure}
\usepackage{url}
\usepackage{todonotes}
\usepackage[hyperfootnotes=false,hidelinks=true]{hyperref}
\usepackage{sansmath}
\usepackage{multirow}
\usepackage[referable]{threeparttablex}
\usepackage{booktabs, longtable}
\newcounter{tablenote}[table]

\usepackage{pifont}
\newcommand{\gap}{-5mm}

\title{Risk and Ambiguity in Information Seeking: Eye Gaze Patterns Reveal
Contextual Behaviour in Dealing with Uncertainty}

\author[1,2]{Peter Wittek}
\author[3,4]{Ying-Hsang Liu}
\author[2]{S\'andor Dar\'anyi}
\author[4]{Tom Gedeon}
\author[5]{Ik Soo Lim}

\affil[1]{ICFO-The Institute of Photonic Sciences\\
Barcelona Institute of Science and Technology\\
08860 Castelldefels (Barcelona), Spain}
\affil[2]{University of Bor{\aa}s\\
50190 Bor{\aa}s, Sweden}
\affil[3]{Charles Sturt University\\
Wagga Wagga NSW 2678, Australia}
\affil[4]{The Australian National University\\
Acton ACT 2601, Australia}
\affil[5]{School of Computer Science\\
Bangor University\\
LL57 1UT Bangor, United Kingdom}

\date{}
\begin{document}
\maketitle
\begin{abstract}
Information foraging connects optimal foraging theory in ecology with how humans search for information. The theory suggests that, following an information scent, the information seeker must optimize the tradeoff between exploration by repeated steps in the search space vs. exploitation, using the resources encountered. We conjecture that this tradeoff characterizes how a user deals with uncertainty and its two aspects, risk and ambiguity in economic theory. Risk is related to the perceived quality of the actually visited patch of information, and can be reduced by exploiting and understanding the patch to a better extent. Ambiguity, on the other hand, is the opportunity cost of having higher quality patches elsewhere in the search space. The aforementioned tradeoff depends on many attributes, including traits of the user: at the two extreme ends of the spectrum, analytic and wholistic searchers employ entirely different strategies. The former type focuses on exploitation first, interspersed with bouts of exploration, whereas the latter type prefers to explore the search space first and consume later. Based on an eye-tracking study of experts' interactions with novel search interfaces  in the biomedical domain, we demonstrate that perceived risk shifts the balance between exploration and exploitation in either type of users, tilting it against vs. in favour of ambiguity minimization. Since the pattern of behaviour in information foraging is quintessentially sequential, risk and ambiguity minimization cannot happen simultaneously, leading to a fundamental limit on how good such a tradeoff can be. This in turn connects information seeking with the emergent field of quantum decision theory.
\end{abstract}

\section{Introduction}
Searching for food is a common pattern of behaviour: humans and animals
share dedicated cognitive mechanisms to find resources in the
environment. Such resources are distributed in spatially localized
patches where the task is to maximize one's intake, that is, knowing
when to exploit a local patch versus when it is time to move on and
explore one's broader surroundings.

In humans, the underlying neuropsychological mechanisms result in
cognitive searches, such as recalling words from
memory~\cite{hills2012optimal,hills2015exploration}. As part of users'
information seeking behaviour, the concept of information foraging
describes the above quest by a similar
strategy~\cite{pirolli1999information}.

Key to the understanding of decisions by a consumer of information is
that they are subject to uncertainty: his or her knowledge of the
environment is incomplete, so the resulting decisions must go back to
perceptions and certain heuristics. By turning to classical works in
economy, we can identify two facets of this uncertainty, namely risk and
ambiguity~\cite{knight1921risk,ellsberg1961risk}. Their interpretation
according to the foraging scenario is in place here.

Briefly, \emph{risk} is the quality of the current patch and our
fragmented perception of it. Is the place of good quality? Should one
stay here or move on? Since we are already at the preselected location,
we do have prior information about it. A risk-minimizing behaviour will
favour exploitation over exploration, staying longer at individual
locations, potentially losing out if outstanding patches remain
unvisited.

The above immediately have anthropological overtones. Foraging behaviour
seems to apply to a much larger domain than just looking for food, such
as the optimization of upper and lower extremities of pleasure and pain,
gain and loss, benefit and cost, reward and punishment, joy and sorrow. Seeking one while avoiding the other is the subject of risk
analysis, where the nature of risk is hesitation. It is obvious that if
we are too quick or too slow, we lose a positive option by gaining a
negative score somewhere else without even having noticed.

\emph{Ambiguity}, on the other hand, is related to opportunity cost, the
price of not foraging elsewhere. ``Elsewhere'' refers to the rest of the
unknown distribution which is not observed at the moment. A human
forager who wants to reduce ambiguity first will jump around different
patches and explore more, learning as much as possible about the
information distribution while reducing the associated uncertainty. This
behaviour will not stop at the first good enough patch.

To continue the anthropological implications, ambiguity would also mean
that all of the above are the essence of situations, of problem solving
in general, but by decisions (and the crucial belief that we have
resolved the problem) we create a new situation by trying to escape it.
So in a sense, risk would belong to the surface layer and ambiguity to
the deep layer of any decision situation. If the above hold, we could
identify many more scenarios relevant from psychology to decision theory
and from cognitive science to the stock exchange.

Resonating with the aforementioned, our working hypothesis below will be
that if animal foraging is subject to uncertainty, and information
seeking is an essentially identical activity in a different context,
then a limit to simultaneous risk and ambiguity minimization must apply
to information foraging as well. This limit emerges from the sequential
and incompatible nature of the decisions made to minimize these two
aspects of uncertainty. The incompatible decisions are similar to measurements in quantum mechanics where they give rise to the uncertainty principle; thus our work connects information foraging and information seeking behaviour to the thriving field of quantum decision theory~\cite{yukalov2008quantum,bruza2009quantum,khrennikov2010ubiquitous,busemeyer2012quantum,ashtiani2015survey}. We will demonstrate our point on eye tracking
studies data in our study of user interactions with novel search
interfaces for biomedical information search.

The structure of the article is as follows: in Section 2 we discuss the
origins and application areas of uncertainty, including foraging
decisions and information seeking as examples. In Section 3, a
preliminary analysis of search behaviour based on eye tracking data is
offered, with Section 4 listing our results and discussing them. Section
5 brings us to our conclusions and plans for future work.

\section{Background}
\subsection{The origins and application areas of uncertainty}
A decision in the presence of uncertainty means that the outcome cannot
be fully predicted before the decision is made. Multiple possible
outcomes can occur, and our knowledge of the probability distribution
only allows for a limited characterization of uncertainty. Following
Refs.~\cite{knight1921risk}
and~\cite{ellsberg1961risk,camerer1992recent}, we can distinguish
between two fundamental aspects of uncertainty, aforementioned ambiguity
and risk. The simple definition of risk is uncertainty with known
probabilities, a certain a priori probability for a given outcome.
Ambiguity is also probabilistic but less well defined, generally
associated with events that the decision maker has even less information
about than the risk of outcomes. The two aspects are also called
expected and unexpected uncertainty. Dealing with unexpected uncertainty
involves a more subjective evaluation of probabilities. In the case of
ambiguity, less information is available, and expected utility is harder
to estimate. Not knowing crucial information, such as the probability
distribution of the outcomes, is a frightening prospect which explains
why most people are ambiguity-averse~\cite{ellsberg1961risk}. The two
forms of uncertainty are so different that dealing with risk and
ambiguity are supported by distinct neural mechanisms in
humans~\cite{huettel2006neural}.

Apart from this probabilistic nature of decisions in an uncertain
environment, there is an even deeper form of uncertainty: the kind we
normally refer to in the context of quantum mechanics. Some
measurements on a quantum system are simply incompatible: measuring one
aspect of the system prevents us from learning more about another aspect
thereof, explored by a different measurement.

As stated by Ref.~\cite{folland1997uncertainty} in what constitutes the
basis of this brief overview, ``There are various mathematical aspects
of the uncertainty principle, including Heisenberg's inequality and its
variants, local uncertainty inequalities, logarithmic uncertainty
inequalities, results relating to Wigner distributions, qualitative
uncertainty principles, theorems on approximate concentration, and
decompositions of phase space''~\cite{folland1997uncertainty}. It is
partly a description of a characteristic feature of quantum mechanical
systems, partly a statement about the limitations of one's ability to
perform measurements on a system without disturbing it, and partly a
meta-theorem in harmonic analysis that can be summed up as follows: ``A
nonzero function and its Fourier transform cannot both be sharply
localized.'' Therefore the principle leads to mathematical formulations
of the physical ideas first developed in Heisenberg's seminal paper of
1927~\cite{heisenberg1927uber}, explored from many angles afterwards.

Initially this rule was carved in stone for a particular case only --
that we cannot simultaneously learn the position and the velocity
(momentum) of a quantum particle with arbitrary precision. Namely
Heisenberg's uncertainty principle states that the standard deviation of
such measurement outcomes on these two complementary aspects of the
system cannot be simultaneously minimized:

\begin{equation} \sigma_X \sigma_P\geq \frac{\hbar}{2}\nonumber
\end{equation} where $\sigma_X$ is the standard deviation on the
measurement of the position, $\sigma_P$ is the standard deviation on the
measurement of the momentum, and $\hbar$ is the reduced Planck's constant or Dirac constant.

This principle was later generalized to arbitrary pairs of incompatible
measurements, and expressed by many other mathematical concepts
different from standard deviation. Incompatible measurements mean that
certain observations on a system do not commute: by making an
observation, we are making a second one in the context created by the
first. In other words, incompability, noncommutativity, and
contextuality are closely related concepts.

Noncommutativity allows the definition of an alternative event algebra
or logic, which in turn leads to applications in decision
theory~\cite{bruza2009quantum,busemeyer2012quantum}. This line of
research is part of a broader trend of applying the mathematical
framework of quantum mechanics in domains outside
physics~\cite{khrennikov2010ubiquitous}.

\subsection{Uncertainty and foraging decisions}
We are especially interested in how risk and ambiguity appear in
sequential decisions. Simultaneous or coordinated decision making, on
the other hand, is more complex, being less common among animals because
it involves comparative evaluation. Pointing at a major difference
between the animal kingdom vs. man, Ref.~\cite{kolling2012neural} showed
that humans are able to choose between these two models in uncertain
environments. A foraging scenario is a good example of sequential
decision making: food resources are available in patches, and a forager
must find an optimal strategy to consume the resources. There is a cost
associated with switching from one patch to another. Uncertainty relates
to the quality of the current patch, the quality of background options
-- the opportunity cost of not foraging elsewhere -- and the environment
is also subject to changes. The forager has to minimize the tradeoff
between exploitation of a patch versus exploration of background
options. The pattern is not restricted to food consumption: for
instance, it pertains to mate selection, retrieving memories, and
consumer decisions. In fact, the same neural mechanism can serve these
different functions~\cite{adams2012neuroethology}.

Optimal foraging theory gives the strategy to follow if the
probabilities can be estimated and updated by the
forager~\cite{macarthur1966optimal,charnov1976optimal}. Ambiguity alters
the behaviour: for example, unexpected forms of uncertainty may trigger
more exploration~\cite{cohen2007should}. We would like to see how
ambiguity and risk can be minimized in sequential decisions, and how
that affects exploration and exploitation.

Many decisions require an exploration of alternatives before committing
to one and exploiting the consequences thereof. This is known as
foraging in animals that face an environment in which food resources are
available in patches: the forager explores the environment looking for
high-quality patches, eventually exploiting a few of them only. The
decisions take place in an uncertain environment: ambiguity about the
quality of patches and the risk of not foraging at better patches force
the forager to accept a tradeoff.

Risk-sensitive foraging is not exclusive to animals, human subjects also
show similar behavioural
patterns~\cite{pietras2003human,rushworth2012valuation}. An optimal
solution between exploration and exploitation is generally not known,
except in cases with strong assumptions about both the environment and
the decision maker~\cite{cohen2007should}. The tradeoff between
exploration and exploitation is also known as the partial-feedback
paradigm, linking the decision model to the description--experience
gap~\cite{hertwig2009description}: people perceive the risk of a rare
event differently if the probability distribution is known (decision
from description) vs. when they have to rely on more uncertain
information (decision from experience).

\subsection{Information seeking as foraging}
To take the next step in our working hypothesis, below we shall look at
a scenario where seeking was exercised by gaze fixation at segments of
user interfaces with significant elements of content, and show that
underlying the seemingly random walks of eye gaze on the screen, there
is order in the patterned data inasmuch as a certain typology of user
behaviour applies to them.

The information foraging nature of the data was recognized by eye
tracking analysis, based on the concept of information scent,
operationalized as ``the proportion of participants who correctly
identified the location of the task answer from looking at upper braches
in the tree" in a study of user interactions with visualization of large
tree structures~\cite{pirolli2000effect}. Ref.~\cite{pirolli2001visual}
provided further theoretical accounts for scanpaths from cognitive
perspectives in which users were able to find information more quickly
when strong information scent was detected. Ref.~\cite{Chi2001Using}
built a computational model for user information needs and search
behaviour based on information scent, and the model and algorithm were
evaluated by simulated studies. More recently, the modeling of user
search behavior using eye tracking techniques has focused on levels of
domain knowledge, user interests, types of search task and relevance
judgments in search
processes~\cite{Cole_2010,Cole_2013,Gwizdka_2014,Vakkari_2014,
Zhang_2015}. However, there is still limited understanding of the effect
of individual differences and user perceptions of search tasks on eye
gaze patterns in information search.
Ref.~\cite{White2016Models-and,White2016Interactio} provided a review of
information foraging and user interactions with search systems.

The eye gaze patterns, an indicator of user attention and cognitive
processes have been extensively studied for designing user interfaces,
such as the functional grouping of interface
menu~\cite{Brumby2015Visua,Goldberg1999Compu}, faceted search
interface~\cite{Kemman2013Eye-T,Kules2009What-} and comparison of
interface layouts~\cite{Kammerer2012Effec}. Information retrieval
researchers have been concerned with users' attention to the ranking
position of documents and different components of search engine results
page~(SERP)~\cite{Cutrell2007What-,Dumais2010Indiv,Kim2016Under,
Lorigo2008Eye-t,Savenkov2011Searc}. These studies generally suggest that
there is no significant difference in users' eye gaze patterns on
comparisons of search interface layouts, and users' attention to
elements of interfaces depends on the length and quality of snippets on
SERPs, as well as the displayed position of search results.

\section{User Experiment}
This study was designed to investigate user gaze and search behaviour in
biomedical search tasks, with particular reference to the user's
attention to and use of the document surrogates (i.e., Medical Subject Headings (MeSH) terms,
title, authors, and abstract). A total of 32 biomedical experts
participated in the controlled user experiment, performing searches on
clinical information for patients. The participants were mostly students
with search engine experience and some academic background in the
biomedical domain.

We used a {\sansmath$4\times4\times2$} factorial design with four search
interfaces, controlled search topic pairs and cognitive styles. A
{\sansmath$4\times4$} Graeco-Latin square design was
used~\cite{Fisher_1935} to arrange the experimental conditions. Each
user was assigned 8 topics in total, with a 7-minute limit for each
topic, and the experiment took about 90 minutes in total.
\subsection{Search interfaces}
Participants searched on four different search interfaces, with a single search
system behind the scenes. The four search interfaces were distinguished by
whether MeSH terms were presented and how the displayed MeSH terms were
generated:

\begin{figure}
\centering
\subfigure[Screenshot of Interface ``B'', suggestions per-query and
  displayed at top.]{
\includegraphics[width=12cm,scale=.6]{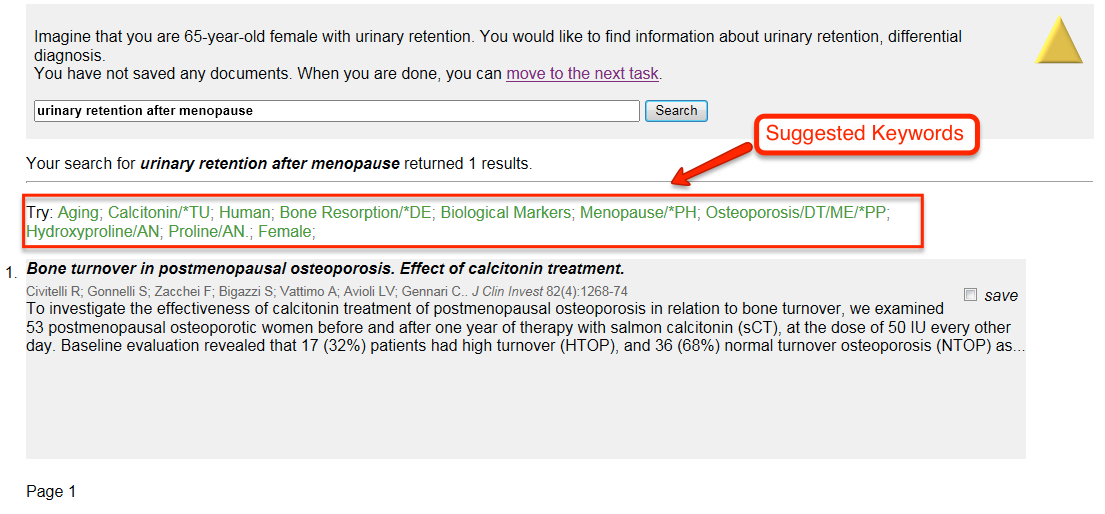}
\label{fig:interface_B}
}\\
\subfigure[Screenshot of Interface ``C'', suggestions per-query and
  displayed at top.]{
\includegraphics[width=12cm,scale=.6]{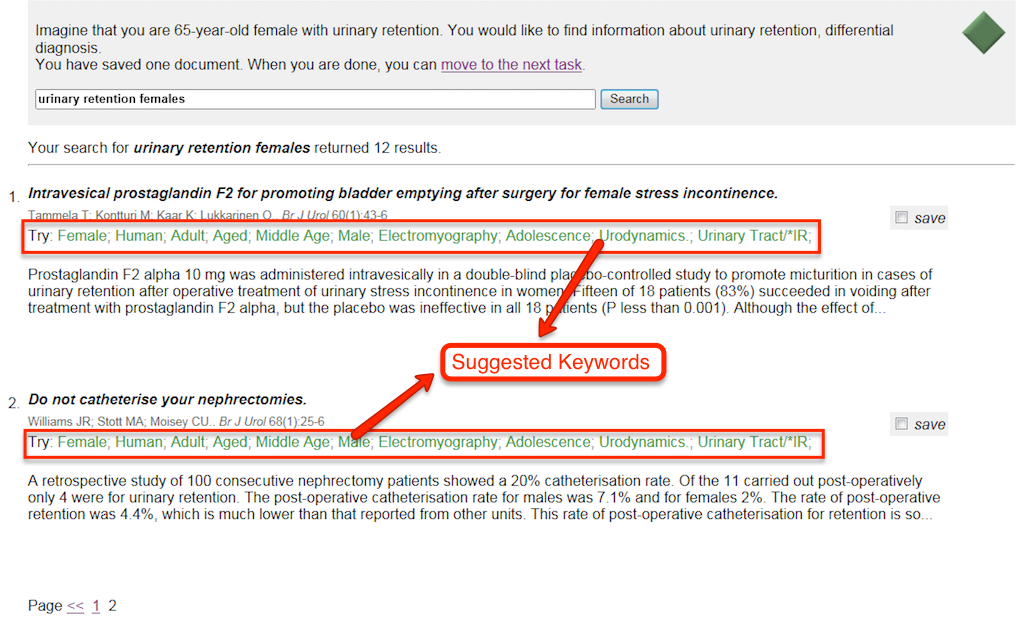}
\label{fig:interface_C}
}\\
\subfigure[Screenshot of Interface ``D'', suggestions per-document and
  displayed with the document.]{
\includegraphics[width=12	cm,scale=.6]{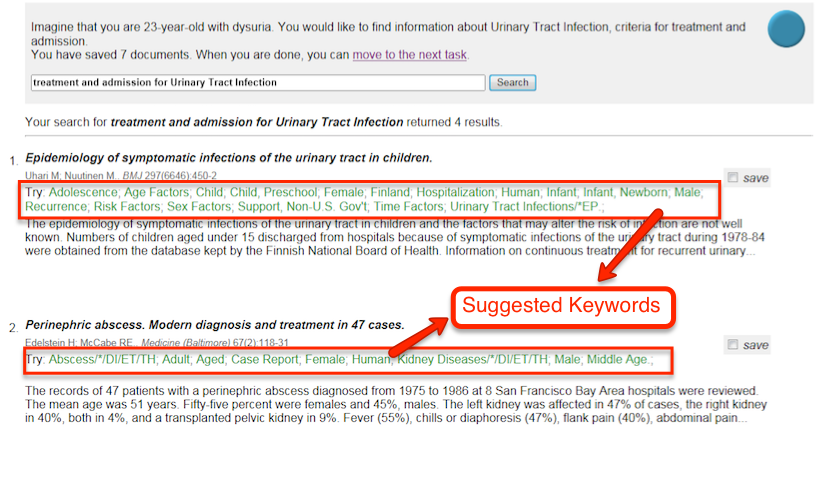}
\label{fig:interface_D}
}
\caption{Three of the four search interfaces in the study.}
\label{fig:three_interfaces}
\end{figure}

\begin{description}

\item[Interface ``A''] mimicked web search and other search systems with
  no controlled vocabulary.  This interface had a brief task
  description at top; a conventional search box and button; and each
  result was represented with its title, authors, publication details,
  and abstract where available.

  Full text was not available, so the results were not clickable.  Users
  judged their success on the titles and abstracts alone.

\item[Interface ``B''] (Figure~\ref{fig:interface_B}) added MeSH terms
  to the interface.  After the user's query was run, MeSH terms from
  all results were collated; the ten most frequent  were displayed at the
  top of the screen.  This mimics the per-query suggestions produced
  by systems like ProQuest\footnote{For example, see
    \url{http://www.proquest.co.uk/en-UK/products/brands/pl_pq.shtml}}.

  MeSH terms were introduced with ``Try:'' and were clickable: if a user
  clicked a term, his or her query was refined to include the MeSH term and
  then re-run.  It was hoped that the label, and the fact they work as
  links, would encourage users to interact with them.

\item[Interface ``C''] (Figure~\ref{fig:interface_C}) used the same MeSH terms as ``B'' but displayed
  them alongside each document, where they may have been more (or less)
  visible.  It is a hybrid of interfaces ``B'' and ``D''.

\item[Interface ``D''] mimicked
  EBSCOhost\footnote{\url{http://www.ebscohost.com/}} and similar
  systems that provide indexing terms alongside each document.  As
  well as the standard elements from interface ``A'', interface ``D''
  displayed the MeSH terms associated with each document, as part of
  that document's surrogate (Figure~\ref{fig:interface_D}).

  Again, terms were introduced with ``Try:'' and were clickable.

\end{description}

Each interface was labelled with a simple figure: a square, circle,
diamond, or triangle, which was referred to in the exit questionnaire. A
save icon alongside each retrieved document was provided to collect user
perceived relevant documents.

\subsection{Search topics}
Search topics used here were a subset of the clinical topics from
OHSUMED~\cite{Hersh_1994}, originally created for information retrieval
system evaluation.  The topics were slightly rewritten so they read as
instructions to the participants (see Figure
\ref{fig:search_topic_example} for an example). Topics were selected to
cover a range of difficulties.

\begin{figure}
\centering \fbox{\parbox[b]{0.9\columnwidth}{Imagine that you are
    63-year-old male with acute renal failure probably 2nd to
    aminoglycosides/contrast dye.\\ You would like to find information
    about acute tubular necrosis due to aminoglycosides, contrast dye,
    outcome and treatment.}}
\caption{An example OHSUMED search topic, reworded for the
  participants.}
\label{fig:search_topic_example}
\end{figure}

\subsection{Procedure}
Participants were given brief instructions about the search task and
system features, followed by a practice topic and then the searches
proper. They were informed that the test collection is incomplete and
out-of-date since the OHSUMED test collection~\cite{Hersh_1994} was
used, with MEDLINE data from 1987 to 1991. User interaction data
recorded included:  all queries, mouse clicks, retrieved and saved
documents, time spent, and eye movements. Electroencephalogram (EEG)
readings were also captured.

Background and exit questionnaires collected demographic information and
asked participants about their perception of the search process.
Participants' opinions of the tasks and the interfaces was sought.
Finally, information on participants' cognitive styles was collected by
a computerised test~\cite{peterson_2003,peterson_2005}, which took a
further 15 minutes to complete.

\subsection{Hardware and software}
The search system was built on
Solr\footnote{\url{http://lucene.apache.org/solr/}}, with the search
results ranked by default relevance score.  The MeSH terms were not
specifically weighted.

Eye gaze data was recorded from two Sony VFCB-EX480B infrared (IR)
cameras which were controlled by Seeing Machines FaceLab~4.5
software\footnote{\url{http://www.seeingmachines.com/product/facelab/}}
and attached to a dedicated machine running Windows~7. At the same time,
EyeWorks Design and EyeWorks
Record\footnote{\url{http://www.eyetracking.com/Software/EyeWorks}} were
used to present instructions for the corresponding search tasks during
the experiment. Gaze points were recorded at 60~Hz, and the eye gaze data
included the $x$ and $y$ coordinates of where the eye was looking on the
screen, as well as the time that gaze point is recorded. EEG data was
recorded with an Emotiv headset\footnote{\url{http://www.emotiv.com/}}
to monitor emotional variation throughout the search session. A
Windows~7 computer was dedicated to the cognitive styles test.

\subsection{Data analysis}
Recordings were analysed to see how often there were fixations in
different parts of document surrogates (i.e., different elements of the
interfaces), and therefore how often people looked at each part.

Four common areas of interest (AOI) were specified: title, author,
abstract and MeSH (except for Interface~A, without MeSH) to investigate
which elements received attention. EyeWorks
Analyze\footnote{\url{http://www.eyetracking.com/Software/EyeWorks}} was
used to specify the AOI, and fixations were specified as gazes within a
5-pixel radius which lasted at least 75~ms~\cite{Marshall_2000}.

In the study a post-search questionnaire was used to assess user
perceptions about the search processes, in which search task difficulty
was also identified as important moderator of eye gaze
behaviour~\cite{Toker_2013}.

\section{Results of search behaviour and eye gaze}
Overall, our results support the hypothesis that search interfaces
have significant effects on eye gaze behaviour in terms of proportion of
fixations in reading time. This in turn translates to different strategies in dealing with risk and ambiguity.

\subsection{Search task difficulty and eye gaze}
Table~\ref{difficulty_aoi}
reveals that there was a statistically significant negative relationship
between user perception of search task difficulty and proportion of
fixations in reading time on all elements in documents. Further analysis
indicates a significant interaction effect of interface and task
difficulty on the fixations time spent in title {($\mathrm {F}(3, 248) =
3.72, p < .05$)} and MeSH terms {($\mathrm {F}(3, 248) = 3.71, p <
.05$)}, but it is not the case for the element of author {($\mathrm
{F}(3, 248) = 1.69, p
> .05$)} and abstract {($\mathrm {F}(3, 248) = 1.55, p > .05$)}.
These results suggest that \emph{perceived the search task as
difficult, they did not attend to all elements in the documents}.

\begin{table}[!htbp] \begin{center} {\footnotesize \scalebox{1.0}{
\begin{tabular}{lrrrrrrrr} \hline & \multicolumn{1}{c}{$Cut Point$} &
\multicolumn{1}{c}{$Odds$} & \multicolumn{1}{c}{$Log$} &
\multicolumn{1}{c}{$Stand.$} & \multicolumn{1}{c}{$t-$} &
\multicolumn{1}{c}{$Stat.$}\\
& \multicolumn{1}{c}{(Mean)} &
\multicolumn{1}{c}{$Ratio$} & \multicolumn{1}{c}{$Odds$} &
\multicolumn{1}{c}{$Error$} & \multicolumn{1}{c}{$Value$} &
\multicolumn{1}{c}{\emph{Signif.}} \\
\hline Areas of Interest &        &        & & & &\\
Title & 24.33  & 0.06   & -2.73  & 0.71   & -3.86  & Yes \\
Author & 12.53 &
0.12   & -2.13  & 0.70   & -3.02  & Yes \\
Abstract & 45.81  & 0.13   & -2.05 &
0.71   & -2.90   & Yes \\
MeSH & 17.34  & 0.07   & -2.72   & 0.70   & -3.87 &
Yes \\
\hline \end{tabular}}}
\end{center}
\vspace{\gap}
\caption[Summary of the
relationship between search task difficulty and eye gaze]{Summary of the
relationship between search task difficulty and eye gaze (N search task
difficulty = 256, N eye gaze = 256; statistical significance at 95~\%)}
\label{difficulty_aoi} \end{table}

\subsection{Search task difficulty, cognitive style and eye gaze}
In the study the E-CSA-WA (Extended Cognitive Style Analysis--Wholistic
Analytic) test was used to determine user's cognitive style. A Wholistic
Analytic ratio (WA ratio) for each participant was
produced~\cite{peterson_2003}. The results suggest that there was no
significant relationship between the users' cognitive style and eye gaze
across all elements in documents in terms of proportion of fixations in
reading time.

Further analysis of the effects of search task difficulty, search
interface and cognitive style and their interactions on eye gaze
indicates significant interaction effects of difficulty and cognitive
style {($\mathrm {F}(1, 240) = 4.54, p < .05$)}, and cognitive style vs.
search interface {($\mathrm {F}(3, 240) = 2.89, p < .05$)} in terms of
fixation time on the element of abstract. We found significant
interaction effects between search task difficulty and search interface
{($\mathrm {F}(3, 240) = 4.19, p < .01$)}, and search interface and
cognitive style {($\mathrm {F}(1, 240) = 4.24, p < .01$)} for the
element of MeSH terms. These results suggest that \emph{searchers with
different cognitive styles may use different search strategies under an
environment with uncertainty perceived as difficult and observed by
their eye gaze behaviour}.

\subsection{Search task difficulty and search behaviour}
Table~\ref{difficulty_search_behaviour} shows that when search tasks
were perceived difficult, users tended to spend less time searching,
issued less queries or typed queries, saved fewer documents and had
fewer mouse clicks, but there was no difference in the number of MeSH
queries issued and the number of pages viewed.

Overall, the results indicate that searchers made less
mental effort when the search tasks were difficult, and they tended to
optimise limited resources in information seeking, demonstrated both by
eye gaze (Table~\ref{difficulty_aoi}) and search behaviour (Table~\ref{difficulty_search_behaviour}). \emph{Search behaviour
associated with expending mental efforts like issuing MeSH terms and
viewing SERPs has not changed according to the uncertainty within the
environment, such as perceived search task difficulty}.

\begin{table}[!htbp] \begin{center} {\footnotesize \scalebox{1.0}{
\begin{tabular}{lrrrrrrrrrrr} \hline & \multicolumn{1}{c}{$Cut Point$} &
\multicolumn{1}{c}{$Odds$} & \multicolumn{1}{c}{$Log$} &
\multicolumn{1}{c}{$Stand.$} & \multicolumn{1}{c}{$t-$} &
\multicolumn{1}{c}{$Stat.$}\\
& \multicolumn{1}{c}{(Mean)} &
\multicolumn{1}{c}{$Ratio$} & \multicolumn{1}{c}{$Odds$} &
\multicolumn{1}{c}{$Error$} & \multicolumn{1}{c}{$Value$} &
\multicolumn{1}{c}{\emph{Signif.}} \\
\hline Search behaviour &        &        & & & &\\
Time spent & 185.7  & 0.11   & -2.25  & 0.70   & -3.20  & Yes \\
Number of queries issued & 3.80 &
0.12   & -2.13  & 0.72   & -2.98  & Yes \\
Number of MeSH queries issued & 0.33  & 0.30   & -1.19 &
0.74   & -1.60   & No \\
Number of typed queries issued & 3.48  & 0.18   & -1.73 &
0.74   & -2.33   & Yes \\
Number of pages viewed & 5.04  & 0.23   & -1.47   & 0.76   & -1.94 &
No \\
Number of saved documents & 3.63  & 0.10   & -2.33   & 0.71   & -3.26 &
Yes  \\
Number of mouse clicks & 4.88  & 0.09   & -2.42   & 0.72   & -3.37 &
Yes \\
\hline
\end{tabular}}}
\end{center}
\vspace{\gap} \caption[Summary of the relationship between search task
difficulty and search behaviour]{Summary of the relationship between
search task difficulty and search behaviour (N search task difficulty =
256, N eye gaze = 256; statistical significance at 95~\%)}
\label{difficulty_search_behaviour}
\end{table}

\subsection{Search behaviour and eye gaze in information search}
\emph{Number of queries issued.} Table~\ref{queries_aoi} shows that
there was a statistically significant connection between the number of
queries issued and the area of interest (AOI) of MeSH terms. That is, when users issued
more queries, they paid they paid significantly more attention.

\begin{table}[!htbp] \begin{center} {\footnotesize \scalebox{1.0}{
\begin{tabular}{lrrrrrrrr} \hline & \multicolumn{1}{c}{$Cut Point$} &
\multicolumn{1}{c}{$Odds$} & \multicolumn{1}{c}{$Log$} &
\multicolumn{1}{c}{$Stand.$} & \multicolumn{1}{c}{$t-$} &
\multicolumn{1}{c}{$Stat.$}\\
  & \multicolumn{1}{c}{(Mean)} &
\multicolumn{1}{c}{$Ratio$} & \multicolumn{1}{c}{$Odds$} &
\multicolumn{1}{c}{$Error$} & \multicolumn{1}{c}{$Value$} &
\multicolumn{1}{c}{\emph{Signif.}} \\
\hline Areas of Interest &        &        & & & &\\
Title & 24.33  & 0.76   & -0.27  & 0.25   & -1.09  & No \\
Author & 12.53 &
0.93   & -0.07  & 0.25   & -0.28  & No \\
Abstract & 45.81  & 0.92   & -0.09 &
0.25   & -0.34   & No \\
MeSH & 17.34  & 1.67   & 0.51   & 0.25   & 2.04 &
Yes \\
\hline \end{tabular}}}
\end{center}
\caption[Summary of the
relationship between number of queries issued and gaze]{Summary of the relationship
between number of queries issued and gaze (N number of queries issued = 256, N eye
gaze = 256; statistical significance at 95~\%)} \label{queries_aoi} \end{table}

\emph{Number of MeSH queries issued.} Table~\ref{mesh_queries_aoi} reveals that there was a statistically significant relation
between the number of MeSH queries issued and the element of abstract negatively. That is, when
users issued more MeSH queries, they paid significantly less attention to the abstract section of documents.

\begin{table}[!htbp] \begin{center} {\footnotesize \scalebox{1.0}{
\begin{tabular}{lrrrrrrrr} \hline & \multicolumn{1}{c}{$Cut Point$} &
\multicolumn{1}{c}{$Odds$} & \multicolumn{1}{c}{$Log$} &
\multicolumn{1}{c}{$Stand.$} & \multicolumn{1}{c}{$t-$} &
\multicolumn{1}{c}{$Stat.$}\\
 & \multicolumn{1}{c}{(Mean)} &
\multicolumn{1}{c}{$Ratio$} & \multicolumn{1}{c}{$Odds$} &
\multicolumn{1}{c}{$Error$} & \multicolumn{1}{c}{$Value$} &
\multicolumn{1}{c}{\emph{Signif.}} \\\hline Areas of Interest &        &        & & & &
\\
Title & 24.33  & 1.39   & 0.33  & 0.37   & 0.89  & No \\
Author & 12.53 &
1.01   & 0.01  & 0.37   & 0.03  & No \\
Abstract & 45.81  & 0.45   & -0.80 &
0.39   & -2.02   & Yes \\
MeSH & 17.34  & 1.77   & 0.57   & 0.38   & 1.50 &
No \\
\hline \end{tabular}}} \end{center} \caption[Summary of the
relationship between number of MeSH queries issued and gaze]{Summary of the relationship
between number of MeSH queries issued and gaze (N number of MeSH queries issued = 256, N eye
gaze = 256; statistical significance at 95~\%)} \label{mesh_queries_aoi} \end{table}

\emph{Number of mouse clicks.} Table~\ref{clicks_aoi} indicates that there was
a statistically significant inverse relationship between the number of
mouse clicks and the title element of documents visited. That is, users
who clicked the mouse more often were less likely to pay attention to
titles.

\begin{table}[!htbp] \begin{center} {\footnotesize \scalebox{1.0}{
\begin{tabular}{lrrrrrrrr} \hline & \multicolumn{1}{c}{$Cut Point$} &
\multicolumn{1}{c}{$Odds$} & \multicolumn{1}{c}{$Log$} &
\multicolumn{1}{c}{$Stand.$} & \multicolumn{1}{c}{$t-$} &
\multicolumn{1}{c}{$Stat.$}\\
$ $  & \multicolumn{1}{c}{(Mean)} &
\multicolumn{1}{c}{$Ratio$} & \multicolumn{1}{c}{$Odds$} &
\multicolumn{1}{c}{$Error$} & \multicolumn{1}{c}{$Value$} &
\multicolumn{1}{c}{\emph{Signif.}} \\\hline Areas of Interest &        &        & & & &
\\
Title & 24.33  & 0.46   & -0.77  & 0.26   & -3.00  & Yes \\
Author & 12.53 &
0.95   & -0.05  & 0.25   & -0.21  & No \\
Abstract & 45.81  & 1.21   & 0.19 &
0.25   & 0.76   & No \\
MeSH & 17.34  & 1.19   & 0.17   & 0.25   & 0.68 &
No \\
\hline \end{tabular}}} \end{center} \caption[Summary of the
relationship between number of mouse clicks and gaze]{Summary of the relationship
between number of mouse clicks and gaze (N number of mouse clicks = 256, N eye
gaze = 256; statistical significance at 95~\%)} \label{clicks_aoi} \end{table}

\emph{Number of pages viewed.} Table~\ref{pages_viewed_aoi} brings
evidence for the same inverse relationship between the number of pages
viewed vs. titles inspected.

\begin{table}[!htbp] \begin{center} {\footnotesize \scalebox{1.0}{
\begin{tabular}{lrrrrrrrr} \hline & \multicolumn{1}{c}{$Cut Point$} &
\multicolumn{1}{c}{$Odds$} & \multicolumn{1}{c}{$Log$} &
\multicolumn{1}{c}{$Stand.$} & \multicolumn{1}{c}{$t-$} &
\multicolumn{1}{c}{$Stat.$}\\
$ $  & \multicolumn{1}{c}{(Mean)} &
\multicolumn{1}{c}{$Ratio$} & \multicolumn{1}{c}{$Odds$} &
\multicolumn{1}{c}{$Error$} & \multicolumn{1}{c}{$Value$} &
\multicolumn{1}{c}{\emph{Signif.}} \\\hline Areas of Interest &        &        & & & &
\\
Title & 24.33  & 0.47   & -0.75  & 0.27   & -2.82  & Yes \\
Author & 12.53 &
0.61   & -0.49  & 0.26   & -1.85  & No \\
Abstract & 45.81  & 1.40   & 0.34 &
0.26   & 1.31   & No \\
MeSH & 17.34  & 1.54   & 0.43   & 0.26   & 1.65 &
No \\
\hline \end{tabular}}} \end{center} \caption[Summary of the
relationship between number of pages viewed and gaze]{Summary of the relationship
between number of pages viewed and gaze (N number of pages viewed = 256, N eye
gaze = 256; statistical significance at 95~\%)} \label{pages_viewed_aoi} \end{table}

\emph{Number of documents saved.} Table~\ref{documents_saved_aoi}
reveals that there was a statistically significant relationship as
regards the number of documents saved vs. abstracts and MeSH terms as
document segments inspected. That is, when users saved more documents,
they paid significantly more attention to the element of abstract, but
less attention to the MeSH.

\begin{table}[!htbp] \begin{center} {\footnotesize \scalebox{1.0}{
\begin{tabular}{lrrrrrrrr} \hline & \multicolumn{1}{c}{$Cut Point$} &
\multicolumn{1}{c}{$Odds$} & \multicolumn{1}{c}{$Log$} &
\multicolumn{1}{c}{$Stand.$} & \multicolumn{1}{c}{$t-$} &
\multicolumn{1}{c}{$Stat.$}\\
  & \multicolumn{1}{c}{(Mean)} &
\multicolumn{1}{c}{$Ratio$} & \multicolumn{1}{c}{$Odds$} &
\multicolumn{1}{c}{$Error$} & \multicolumn{1}{c}{$Value$} &
\multicolumn{1}{c}{\emph{Signif.}} \\\hline Areas of Interest &        &        & & & &
\\
Title & 24.33  & 1.09   & 0.08  & 0.25   & 0.32  & No \\
Author & 12.53 &
1.32   & 0.28  & 0.26   & 1.08  & No \\
Abstract & 45.81  & 1.72   & 0.54 &
0.26   & 2.10   & Yes \\
MeSH & 17.34  & 0.38   & -0.97   & 0.26   & -3.70 &
Yes \\
\hline \end{tabular}}} \end{center} \caption[Summary of the
relationship between number of documents saved and gaze]{Summary of the relationship
between number of documents saved and gaze (N number of documents saved = 256, N eye
gaze = 256; statistical significance at 95~\%)} \label{documents_saved_aoi} \end{table}

\emph{Summary of search behaviour and gaze pattern types}
Table~\ref{tab:search_behaviour_gaze_patterns} provides a summary
of search behaviours and gaze patterns. These results clearly show that types of
searching behaviour, such as issuing queries with MeSH terms that imply
notable mental effort and strive at the exploitation of resources, are
correlated with changes in eye gaze patterns.

\begin{table*}[!htbp] \centering \caption[Summary of the relationship
between search behaviour and gaze patterns]{Summary of the relationship
between search behaviour and gaze patterns} \vspace{.5em}
\resizebox{\columnwidth}{!}{%
\begin{threeparttable} \scalebox{1.0}{
\begin{tabular}{lccccccc} \hline &
\multirow{2}{*}{\begin{tabular}[c]{@{}c@{}}\# of queries\\
issued\end{tabular}} &
\multirow{2}{*}{\begin{tabular}[c]{@{}c@{}}\# of MeSH queries\\
issued\end{tabular}} &
\multirow{2}{*}{\begin{tabular}[c]{@{}c@{}}\# of mouse\\
clicks\end{tabular}} &
\multirow{2}{*}{\begin{tabular}[c]{@{}c@{}}\# of pages\\
viewed\end{tabular}} &
\multirow{2}{*}{\begin{tabular}[c]{@{}c@{}}\# of documents\\
saved\end{tabular}} \\
\\
\hline
Title    &  \textemdash     &   \textemdash    &  \ding{109}
& \ding{109}    &  \textemdash              \\
Author   &  \textemdash     &   \textemdash   &
\textemdash   &  \textemdash    &   \textemdash             \\
Abstract &  \textemdash     &
\ding{109}    &  \textemdash   & \textemdash     &  \ding{108}      \\
MeSH     & \ding{108}     &
\textemdash   & \textemdash    & \textemdash     & \ding{109}
\\
\hline \end{tabular}}
\begin{tablenotes} \small \item Note. The relationship is not statistically
significant (\textemdash), positively significant (\ding{108}), or negatively
significant (\ding{109}). \end{tablenotes} \end{threeparttable}}
\label{tab:search_behaviour_gaze_patterns} \end{table*}

\section{Results and discussion}
We summarize the main findings in the data as follows:
\begin{itemize}
\item When users perceived their search tasks as difficult, they did not
attend to all content elements in documents.
\item Searchers with different cognitive styles may use different search
strategies under an environment with uncertainty they perceive as
difficult.
\item Search behaviour associated with expanding mental efforts like
issuing MeSH terms and viewing SERPs has not changed according to the
uncertainty within the environment, such as perceived search task
difficulty.
\item Certain search behaviour types, such as issuing queries and MeSH
terms that involve notable mental efforts and exploitation of resources,
are correlated with changes in eye gaze patterns.
\end{itemize}

These findings indicate distinct strategies in dealing with uncertainty,
possibly changing from preferring exploration to exploitation and vice versa, and
therefore corroborate our hypothesis that the corresponding observations
do not commute (Section~\ref{diff}). This in turn enables us to frame
information foraging as a form of quantum-like behaviour
(Section~\ref{quant}).

\subsection{Different strategies and noncommuting observations}\label{diff}
In the above eye tracking study, the document surrogates and the four
layouts characterize different perceptions of risk of information
patches, gazing time being a good figure of merit for exploitation.
Exploration is the jumping gaze combined with a repeated query as these
reduce overall ambiguity. There is evidence that wholistic users prefer
to get an overview of tasks before drilling down to detail, whereas
analytic users look for specific information. These two extreme user
behaviours rely on the two measurement operators, namely risk- vs.
ambiguity reduction, in different order, proving noncommutativity.
Unfortunately, at this point there is no significant relationship
between the users' cognitive style and the AOIs.

However, if we also change the perceived risk by varying the search
interface, the picture changes. The effect of cognitive styles,
interfaces and their interactions on the AOI of MeSH terms (excluding
Interface A) is statistically significant in terms of cognitive style
and interface interactions, and weakly significant in terms of cognitive
style (F(1,188) = 2.79, $p < .01$). Interfaces make a statistically
significant difference for the wholistic style (F(2, 111) = 6.58, $p <
.001$), and cognitive styles make a statistically significant difference
in Interface B (F(1, 62) = 5.11, $p < .05$). The results indicate that
wholistic users' attention to the MeSH terms is more affected by search
interfaces than that of analytic users, and this interaction effect is
significant when interacting with Interface B. Thus noncommutative
measurements emerge.

\subsection{Information seeking is quantum-like}\label{quant}
To sum up, uncertainty as a composite of risk and ambiguity drives
information seeking behaviour in a complex way, with successive
decisions attempting to minimize both components at the same time.
However, to find their joint optimum is not possible, because risk-prone
and ambiguity-prone behaviour manifest two versions of foraging
attitude, called the ``consume first and worry later'' (exploitation)
vs. the ``worry first and consume later'' (exploration) types. Whichever
option taken, it becomes the context of the opposite alternative, so
that ambiguity minimization dependent on risk minimization vs. risk
minimization dependent on ambiguity minimization yield different sets of
retrieved items, i.e. the outcome of information seeking as a process is
non-commutative.

For every case where this joint optimum seeking mentality influences the
results, plus the decision making process that has led to a particular
outcome must be preserved for future reconstruction, our findings are
relevant. However, there is more to the implications of the above.

In this study we have seen that two types of information seeking
behaviour emerged from interaction between the cognitive apparatus and
the phenomenon observed, i.e. information. This is reminiscent of the
the Copenhagen interpretation of quantum mechanics, where interaction between the
measurement apparatus and the observable cannot be reduced to zero, and
the measured value is a result of (or is not independant from)
interaction; again in the words of Ref.~\cite{folland1997uncertainty},
``the values of a pair of canonically conjugate observables such as
position and momentum cannot both be precisely determined in any quantum
state.'' Further, we have found that the above two types of behaviour go
back to the application of two operators, risk- and ambiguity-aversion,
so that by applying now this, then the other first, their sequential
application leads to different results, called non-commutativity.

Moreover, as much as risk and ambiguity are two sides of the same coin,
non-commutativity is an essential feature of the uncertainty principle
core to quantum mechanics. Given this, our current finding hints at something
potentially fundamental about the nature of browsing. At the same time,
since Ref.~\cite{dominich2001mfi} proposed to treat precision and recall
as complementary operators regulating the surface of effectiveness in
information retrieval, whereas Ref.~\cite{vanrijsbergen04:geom} argued
that relevance is an operator on Hilbert space and as such is part of
the quantum measurement process, neither was our insight totally unexpected.
Rather, connected to the uncertainty principle, we see noncommuting
measurements to surface also in information seeking as another link to
quantum decision
theory~\cite{wittek2013foraging,ashtiani2014contextuality,aerts2016from}.

\section{Conclusions and future research}
We interpreted risk and ambiguity as two types of measurement on an
uncertain environment, arguing that in an information foraging scenario,
these measurements are sequential and do not commute, that is, reversing
their order yields different outcomes. We demonstrated this by analyzing
user behaviour in interacting with different designs of search results,
specifically, by tracking the gaze of users. Depending on the degree of
uncertainty involved, qualitatively different types of information
seeking behaviour emerged, agreeing with our hypothesis.

We have reason to believe that similar data, such as clickstreams, will
show similar patternedness as evidence of non-commutative user behaviour
manifesting the same cognitive types in a different setting. In a
broader context, noncommuting measurements are standard tools in quantum
mechanics, and they are being explored in quantum decision theory for
modelling decision problems and known fallacies -- our work connects
information seeking to this line of research.

\section{Acknowledgements}
Peter Wittek and S\'andor Dar\'anyi were supported by the European
Commission Seventh Framework Programme under Grant Agreement Number
FP7-601138 PERICLES. This study was in part funded by the 2014 ALIA
(Australian Library and Information Association) Research Grant Award.
Dr Paul Thomas, Jan-Felix Schmakeit, Marijana Bacic and Xindi Li's
assistance in user experiment are acknowledged. The views and opinions
expressed in this article are those of the authors and do not
necessarily reflect the official policy or position of ALIA or any
author affiliated organisation.


\end{document}